\documentclass{ws-procs975x65}

\begin{document}

\title{Metastable D-term Dynamical SUSY Breaking}

\author{Nobuhito Maru}

\address{Department of Physics and Research and Education Center for Natural Sciences, \\Keio University,
Yokohama, 223-8521, Japan\\
}

\begin{abstract}
We present the mechanism of the dynamical supersymmetry (SUSY) breaking at the metastable vacuum 
in the ${\cal N}=1$ $U(N)$ SUSY gauge theory with adjoint superfields. 
The dynamical SUSY breaking is triggered by the non-vanishing $D$-term coupled to the observable sector, 
and is realized by the self-consistent Hartree-Fock approximation of the NJL type 
while it eventually brings us the non-vanishing $F$-term as well. 
We numerically check the local stability of our metastable vacuum. 
\end{abstract}


\bodymatter

\section{Introduction}
Spontaneous breaking of SUSY occurs much less frequent compared with that of internal symmetry
 in quantum field theory and has attracted much interest \cite{Witten1} of theorists for over the three decades.
Mass hierarchy in elementary particle physics indicates that it is most desirable to break ${\cal N}=1$ SUSY dynamically. 
In fact, the non-renormalization theorem \cite{GRS} protects the generation of holomorphic operator in perturbation theory 
and instanton generated nonperturbative superpotentials have been the major source of dynamical SUSY breaking (DSB).

We focus our attention on general ${\cal N}=1$ theory in four dimensions 
consisting of vector superfields and chiral superfields in the adjoint representation 
with a non-canonical gauge kinetic function. 
It has recently been shown in refs. \cite{imaru} that, in this general situation, 
SUSY is dynamically broken in the metastable vacuum. 
The mechanism that triggers the DSB is the condensate of the Dirac bilinear, 
forcing one of the order parameters $D$ of SUSY to be non-vanishing. 
This is very much reminiscent of the Nambu-Jona Lasinio (NJL) theory \cite{NJL} of broken chiral symmetry 
and hence the BCS superconductivity \cite{BCS}, 
being formulated in terms of the effective action of the auxiliary field whose stationary value is the order parameter.
The method of approximation employed is the self-consistent Hartree-Fock (HF) approximation 
where the tree and the one-loop contributions are regarded as comparable.
Once this mechanism operates, non-vanishing $F$-term is shown to be induced and contributes, 
for instance, to the mass of the fermions. 
The mechanism requires massive adjoint scalars, in particular, the scalar gluons and, 
together with the feature that the D term triggers the breaking, is quite distinct from the previous proposals \cite{ADS, Veneetal} of DSB 
both from theoretical and experimental perspectives. 
The overall $U(1)$ where the non-vanishing $D$ and the Nambu-Goldstone fermion (NGF) reside serves as the hidden sector 
and no messenger field is necessary \cite{imaru} as non-vanishing third prepotential derivatives connect the $U(1)$ sector 
with the observable $SU(N)$ sector \cite{FIS}.

In the next section, 
we review the original reasoning that has led us to the D-term triggered DSB. 
We set up the background field formalism to be used in the subsequent sections, 
separating the three kinds of background from the fluctuations. 
In section three, we elaborate upon our treatment of the effective potential with the three kinds of background fields 
as well as the point of the HF approximation in refs. \refcite{imaru}. 
In section four, 
we present a qualitative argument of our variational analyses of the effective potential. 
Treating $F$-term as an induced perturbation, 
we demonstrate that the stationary values $(D_*, \varphi_*, \bar{\varphi_*})$ are determined by the intersection of the two real curves.  
Numerical analysis is provided that demonstrates the existence of such solution as well as the self-consistency of our analysis. 
The second variation of the scalar potential is computed and the local stability of the vacuum is shown from the numerical data. 
Summary is given in the last section.  
   
\section{The action, assumptions and some properties}

The lagrangian we consider is 
\begin{eqnarray}
    {\cal L}
     &=&     
           \int d^4 \theta K(\Phi^a, \bar{\Phi}^a) + (gauging)   
        + \int d^2 \theta
           {\rm Im} \frac{1}{2} 
           \tau_{ab}(\Phi^a)
           {\cal W}^{\alpha a} {\cal W}^b_{\alpha}
            \nonumber \\
            &&+ \left(\int d^2 \theta W(\Phi^a)
         + c.c. \right).       
           \label{KtauW}
    \end{eqnarray}
At the lowest order in perturbation theory, 
there is no source which gives a vacuum expectation value (vev) to the auxiliary field $D^0$: $\langle D^0 \rangle_{{\rm tree}} = 0$. 
The $U(N)$ gaugino is massless at the tree level while the fermionic partner of the scalar gluon receives 
 the tree level mass $m_a = m_0 = \langle g^{00} \partial_0 \partial_0 W \rangle_{{\rm tree}} $. 

It is useful to summarize here a set of assumptions. 
\begin{enumerate} 
\item a general ${\cal N}=1$ SUSY action 
of chiral superfield $\Phi^a$ in the adjoint representation 
and the vector superfield $V^a$ with the K\"{a}hler potential $K(\Phi^a, \bar{\Phi}^a)$ with its gauging, 
the gauge kinetic superfield $\tau_{ab}(\Phi^a)$ following from the second derivatives of a generic holomorphic function ${\cal F}(\Phi^a)$, 
and the superpotential $W(\Phi^a)$.  
\item third derivatives of ${\cal F}(\Phi^a)$ at the scalar vev's are non-vanishing. 
\item the superpotential preserves ${\cal N}=1$ SUSY at tree level. 
\item the gauge group is $U(N)$ and the vacuum is taken to be in the unbroken phase of $U(N)$. 
\end{enumerate}
In refs. \refcite{imaru}, it was shown that  the vacuum develops a non-vanishing vev of an auxiliary field $D^0$ in the HF approximation. 
The theory, therefore, realizes the D-term DSB. 
The relatively simple estimate has shown that the vacuum can be made long lived.  
Let us recall a few more key aspects.

The part of the lagrangian which produces the fermion mass matrix of size $2N$ 
is
\begin{eqnarray}
-\frac{1}{2} (\lambda^a, \psi^a) 
\left(
\begin{array}{cc}
0 & -\frac{\sqrt{2}}{4} {\cal F}_{abc} D^b \\
-\frac{\sqrt{2}}{4} {\cal F}_{abc} D^b & \partial_a \partial_c W \\
\end{array}
\right) 
\left(
\begin{array}{c}
\lambda^c \\
\psi^c \\
\end{array}
\right) + (c.c.).
\label{massterm}
\end{eqnarray} 
 It was observed that the auxiliary $D^a$ field,
  which is an order parameter of ${\cal N} =1$ SUSY,
    couples to the fermionic (but not bosonic) bilinears through the third prepotential derivatives:
     the non-vanishing vev of $D^0$ immediately gives a Dirac mass to the fermions.
  Equation of motion for $D$-term implies
  \begin{eqnarray}
    \langle D^{0} \rangle
    =    - \frac{1}{2 \sqrt{2}} \langle g^{00} 
           \left( {\cal F}_{0cd}\psi^d \lambda^c 
         + \bar{{\cal F}}_{0cd} \bar{\psi}^d \bar{\lambda}^c \right) \rangle,
 \end{eqnarray}
 telling us that the condensation of the Dirac bilinear is responsible for $\langle D^{0} \rangle \neq 0$.
 
 We diagonalize the holomorphic part of the mass matrix: 
\begin{eqnarray}
  M_{F a} \equiv
\left(
\begin{array}{cc}
0 & -\frac{\sqrt{2}}{4} \langle {\cal F}_{0aa} D^0 \rangle \\
-\frac{\sqrt{2}}{4} \langle {\cal F}_{0aa} D^0 \rangle & \langle \partial_a \partial_a W \rangle \\
\end{array}
\right). 
\label{massmatrix1}
\end{eqnarray} 
 Note that the non-vanishing third prepotential derivatives are ${\cal F}_{0aa}$
 where $a$ refers to the generators of the unbroken gauge group.
The two eigenvalues of eq. (\ref{massmatrix1}) for each generator are 
 \begin{eqnarray}
 \label{eigenvalue}
 {\Lambda}_{a {\bf 11}}^{(\pm)} =  
 \langle \partial_a \partial_a W \rangle  \lambda_{a{\bf 11}}^{(\pm)}, \quad 
 \lambda_{a{\bf 11}}^{(\pm)} \equiv \frac{1}{2}\left( 1 \pm \sqrt{1 +\Delta_{{\bf 11}}^2} \right), \quad
  \Delta_{a{\bf 11}}^2 \equiv   \frac{\langle {\cal F}_{0aa} D^0 \rangle^2}{2 \langle \partial_a \partial_a W \rangle^2}. 
  \label{lambdapm}
\end{eqnarray}
It was also shown in refs. \refcite{imaru} that the non-vanishing $F^0$ term is induced
 by the consistency of our procedure of computation. (See also ref. \refcite{supersoft}). 
  This is because the stationary value of the scalar fields gets shifted upon the variation (the vacuum condition). 
  The final mass formula for the $SU(N)$ fermions is to be read off from  
\begin{eqnarray}
{\cal L}_{{\rm mass}}^{(holo)} &=& -\frac{1}{2} \langle g_{0a,a} \rangle \langle \bar{F}^0 \rangle \psi^a \psi^a 
+ \frac{i}{4} \langle {\cal F}_{0aa} \rangle \langle F^0 \rangle \lambda^a \lambda^a
-\frac{1}{2} \langle \partial_a \partial_a W \rangle \psi^a \psi^a \nonumber \\
&&+ \frac{\sqrt{2}}{4}
\langle {\cal F}_{0aa} \rangle \psi^a \lambda^a 
\langle D^0 \rangle. 
\end{eqnarray}
The mass matrix is read off as
\begin{eqnarray}
{\cal M}_a &=&
\left(
\begin{array}{cc}
-\frac{i}{2} g^{aa} {\cal F}_{0aa} F^0, & -\frac{\sqrt{2}}{4} \sqrt{g^{aa} ({\rm Im}{\cal F})^{aa}}{\cal F}_{0aa} D^0 \\
-\frac{\sqrt{2}}{4} \sqrt{g^{aa} ({\rm Im}{\cal F})^{aa}}{\cal F}_{0aa} D^0, & g^{aa} \partial_a \partial_a W + g^{aa} g_{0a,a} \bar{F}^0 \\
\end{array}
\right) 
= \left(
\begin{array}{cc}
m_{\lambda\lambda}^a & m_{\lambda\psi}^a \\
m_{\psi \lambda}^a  
& m_{\psi\psi}^a \\
\end{array}
\right). \nonumber \\
\label{imaru3massmatrix}
\end{eqnarray}
We parametrize this matrix such that, in the case of $F^0=\bar{F}^0=0$, its form reduces to that of refs. \refcite{imaru}.
The quantities with multiple indices such as ${\cal F}_{0aa}$ receive $U(N)$ invariant expectation values:
$\langle {\cal F}_{0aa} \rangle = \langle {\cal F}_{000} \rangle$ e.t.c.
We suppress the indices as we work with the unbroken $U(N)$ phase in this paper.
The two eigenvalues of the holomorphic mass matrix are written as
\begin{eqnarray}
\Lambda^{(\pm)} \equiv ({\rm tr}{\cal M}) \lambda^{(\pm)}, \quad
\label{ev1}
\lambda^{(\pm)} = \frac{1}{2} \left( 1 \pm \sqrt{(1+if)^2 + \left( 1+\frac{i}{2}f \right)^2 \Delta^2 } \right)
\label{ev2}
\end{eqnarray}
where $\Delta \equiv -  2m_{\lambda\psi}/m_{\psi\psi},~f \equiv 2im_{\lambda\lambda}/{\rm tr}{\cal M}$.

\section{The effective potential in the HF approximation}


In the HF approximation, one begins with considering the situation 
where one-loop corrections in the original expansion in $\hbar$ become large 
and are comparable to the tree contribution. 
In this section, we start the analysis of this kind for our effective potential. 
There are three constant background fields as arguments of the effective potential: 
$\varphi \equiv \varphi^0~({\rm complex})$, $U(N)$ invariant background scalar, 
$D \equiv D^0~({\rm real})$ and $F \equiv F^0~({\rm complex})$. 
The latter two are the order parameters of ${\cal N}=1$ SUSY. 

Let us denote our effective potential by $V$. 
\begin{eqnarray}
V = V_{{\rm tree}} + V_{{\rm c.t.}} + V_{{\rm 1-loop}}. 
\end{eqnarray}
The first term is the tree contributions, the second one is the SUSY counterterm and the last one is the one-loop contributions, 
which are explicitly given by 
\begin{eqnarray} 
&&V_{{\rm tree}} = - gF\bar{F} -\frac{1}{2} ({\rm Im}{\cal F}'') D^2 - F W' - \bar{F}\bar{W'},
\label{tree} \\
&&V_{{\rm c.t.}} = -\frac{1}{2} {\rm Im} \int d^2\theta \Lambda {\cal W}^{0\alpha} {\cal W}_{0\alpha} 
= -\frac{1}{2} ({\rm Im} \Lambda) D^2, \\ 
&&V_{{\rm 1-loop}} = \frac{N^2|{\rm tr}{\cal M}|^4}{32\pi^2} 
\left[
A(\varepsilon, \gamma) 
\left(
|\lambda^{(+)}|^4 + |\lambda^{(-)}|^4 - \left|\frac{m_s}{{\rm tr}{\cal M}} \right|^4
\right) \right. \nonumber \\
&& \left. 
-|\lambda^{(+)}|^4\log |\lambda^{(+)}|^2 
-|\lambda^{(-)}|^4\log |\lambda^{(-)}|^2
+\left| \frac{m_s}{{\rm tr}{\cal M}} \right|^4 \log \left| \frac{m_s}{{\rm tr}{\cal M}} \right|^4
\right]
\end{eqnarray}
where the scalar gluon mass is $m_s(\varphi, \bar{\varphi}) \equiv g^{-1}(\varphi, \bar{\varphi}) W''(\varphi)$. 
The one-loop potential is calculated by the dimensional reduction scheme and the divergence reside in 
\begin{eqnarray}
A(\varepsilon, \gamma) = \frac{1}{2} -\gamma +\frac{1}{\varepsilon}, \qquad \varepsilon = 2-\frac{d}{2}. 
\end{eqnarray}
The divergence is subtracted by a counterterm associated with ${\rm Im}{\cal F}''$ by setting up a renormalization condition
\begin{eqnarray}
\left. \frac{1}{N^2} \frac{\partial^2 V}{(\partial D)^2}
 \right|_{D=0, \varphi = \varphi_*, \bar{\varphi} = \bar{\varphi}_* } =2c. 
\label{rencond}
\end{eqnarray} 
Note that this condition is set up at $D=0$ and the stationary point of the scalar.

\section{Stationary conditions, gap equation and stability of our false vacuum}

Now we turn to our variational problem. 
It is stated as 
\begin{eqnarray}
&&\frac{\partial V}{\partial D} = 0, \label{Dflat} \\
&&\frac{\partial V}{\partial F} = 0~{\rm and~its~complex~conjugate}, \label{Fflat}\\
&&\frac{\partial V}{\partial \varphi} = 0~{\rm and~its~complex~conjugate}. \label{phiflat}
\end{eqnarray}
We will regard the solution to be obtained by considering eqs. (\ref{Dflat}) and (\ref{phiflat}) first and 
solving $D$ and $\varphi$ for $F$ and $\bar{F}$:
\begin{eqnarray}
D=D_*(F, \bar{F}), \quad \varphi = \varphi_*(F, \bar{F}), \quad \bar{\varphi} = \bar{\varphi}_*(F, \bar{F}).
\end{eqnarray}
Eq. (\ref{Fflat}) is then
\begin{eqnarray}
\left. 
\frac{\partial V(D=D_*(F, \bar{F}), \varphi = \varphi_*(F, \bar{F}), \bar{\varphi} = \bar{\varphi}_*(F, \bar{F}), F, \bar{F})}{\partial F}
\right|_{D, \varphi, \bar{\varphi}, \bar{F}~{\rm fixed}} =0
\end{eqnarray}
and its complex conjugate. 
These will determine $F=F_*, \bar{F}=\bar{F}_*$. 

Here, we are going to work in the region
where the strength $\left| F_* \right|$ is small and can be treated perturbatively. 
This means that, in the leading order, the problem posed by eq. (\ref{Dflat}) and eq. (\ref{phiflat}) becomes
\begin{eqnarray}
&&\frac{\partial V(D, \varphi, \bar{\varphi}, F=0, \bar{F}=0)}{\partial D} = 0, \label{Dflatleading} \\
&&\frac{\partial V(D, \varphi, \bar{\varphi}, F=0, \bar{F}=0)}{\partial \varphi} =  \frac{\partial V(D, \varphi, \bar{\varphi}, F=0, \bar{F}=0)}{\partial \bar{\varphi}} = 0
\label{phiflatleading}
\end{eqnarray}
and this problem does not involve the tree potential eq. (\ref{tree}) except the $D^2$ term,
 as $F$ and $\bar{F}$ are set zero. 
Eq. (\ref{Dflatleading}) is nothing but the gap equation, 
while eq. (\ref{phiflatleading}) is the stationary conditions for the scalar.  
This is the variational problem which we analyze. 
A set of stationary values $(D_*, \varphi_*, \bar{\varphi}_*)$ is determined as the solution. 

In this paper, we do not write down the explicit expressions for the gap equation and the stationary conditions for the adjoint scalar 
because of their complicated forms (See the last paper in refs. \refcite{imaru} for details) and the limitation of the space to describe. 
Instead,  we give a qualitative argument how these conditions are solved and numerical results in the simplified case. 

\begin{figure}[htbp]
 \begin{center}
  \includegraphics[width=40mm]{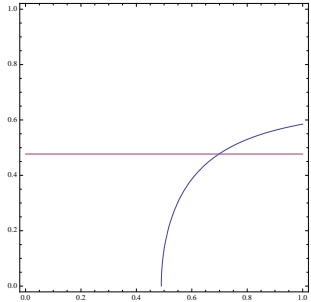}
 \end{center}
 \caption{The schematic picture of the intersection of the two curves which represent 
 the solution to the gap equation (the red one) 
 and the $\varphi$ flat condition (the blue one). The horizontal axis is denoted by $\varphi/M$ and 
the vertical one by $\Delta_0$. 
 The values at the stationary point ($\Delta_{0*}, \varphi_*=\bar{\varphi}_*$) are
  read off from the intersection point. }
 \label{intersection}
\end{figure}

The stationary values ($D_*, \varphi_*, \bar{\varphi}_*$) are determined by 
the gap equation and the stationary condition. 
The solution to the stationary condition in the $\Delta_0$ profile is determined as the point of intersection of the potential 
with the quadratic term having $\varphi=\bar{\varphi}$ dependent coefficients. 
Actually, it is a real curve in the full ($\Delta_0, \varphi=\bar{\varphi}$) plane. 
Likewise, the solution to the gap equation provides us with another real curve in the  ($\Delta_0, \varphi=\bar{\varphi}$) plane. 
The values ($\Delta_{0*}, \varphi_*=\bar{\varphi}_*$) are the intersection of these two. 
The schematic figure of the intersection is displayed in Figure \ref{intersection}. 
By tuning our original input functions, it is possible to arrange such intersection. 

We study some numerical solutions to the gap equation 
and the stationary condition for $\varphi$ 
in the real $\Delta_0$ case. 
In order to find $\varphi_*$ explicitly, the forms of the prepotential ${\cal F}$ and the superpotential $W$ 
must be specified. 
Here, we take a simple prepotential and a superpotential as 
\begin{eqnarray}
{\cal F} = \frac{i}{2N}  {\rm tr} \varphi^2 + \frac{1}{3!MN} {\rm tr} \varphi^3,  \quad 
\label{F} 
W = \frac{m^2}{N} {\rm tr} \varphi + \frac{1}{3!N} {\rm tr} \varphi^3, 
\label{W}
\end{eqnarray}
where 
$m, M$ carry dimensions. 
In particular, $M$ is a cutoff scale of the theory.

\begin{figure}[htbp]
 \begin{center}
  \includegraphics[width=80mm]{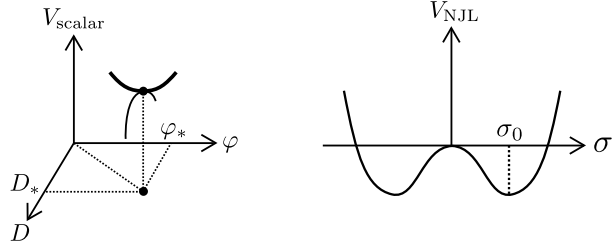}
 \end{center}
 \caption{Comparison of $V_{{\rm scalar}}$ around the stationary value $(D_*, \varphi_*)$ with $V_{{\rm NJL}}$.} 
 \label{potential}
\end{figure}

Some numerical solutions to the gap equation and the stationary condition for $\varphi$ 
in some parameter points are listed in the Table below. 
In these examples, we have taken some values of $-\frac{N^2}{{\rm Im}(i + \Lambda)}$ and $m$ just for an illustration and 
the ratio $|F_*/D_*|$ is evaluated. 
We can find that the $F$-term is smaller than the $D$-term in these examples. 
Also, the scalar gluon mass squared are positive\footnote{In the Table, only the leading term of the scalar gluon mass is shown. 
See the last paper in refs. \refcite{imaru} for more details.}, which ensures a local stability of our vacuum. 
\begin{table}[htb]
\begin{center}
  \begin{tabular}{|c|c|c|c|} 
  \hline
$\Delta_{0*}$  
& $\varphi_*/M~(-\frac{N^2}{{\rm Im}(i+\Lambda)})$ & $|F_*/D_*|$ & $m^2_\varphi$ \\ 
\hline
0.477 & 0.707~(10000) & 0.524~($m \ll M$) & 0.4998 \\
\hline
1.3623 & 0.8639~(2000) & 0.224~($m \ll M$) & 0.7463  \\
1.3623 & 0.5464~(5000) & 0.142~($m \ll M$) & 0.2986 \\
1.3623 & 0.3863~(10000) & 0.100~($m \ll M$) & 0.1492 \\
 \hline
  \end{tabular}
\end{center}
\caption{Samples of numerical solutions for the gap equation and the stationary condition for $\varphi$. 
The ratio $|F_*/D_*|$ and $|f_{3*}|$ are also evaluated for consistency check.}
\label{numerical}
\end{table}
As a summary of our understanding, 
a schematic figure is drawn in Fig. \ref{potential}, which illustrates the local stability of the scalar potential 
at the vacuum of dynamically broken SUSY in comparison with the well-known NJL potential.

\section{Summary}

We have proposed a new mechanism of DSB at the metastable vacuum 
in a ${\cal N}=1$ SUSY $U(N)$ gauge theory with an adjoint chiral super field and non-canonical gauge kinetic term, 
in which non-vanishing $D$-term vev is developed in the self-consistent HF approximation of NJL type. 
Non-zero $F$-term is also induced by such a $D$-term as well. 
We analyzed a gap equation for $D$-term and the stationary conditions for $\varphi$ and $F$-term 
in the case that $F$-term can be treated as a perturbation comparing to $D$-term. 
The solutions are obtained as the intersection point of two curves 
representing the gap equation and the stationary conditions. 
Numerical examples of the solutions are found in the simplified case 
and the local stability of vacua we found is also confirmed by the numerical data. 

\section*{Acknowledgements}
The work of the author is supported in part by the Grant-in-Aid for Scientific Research  
  from the Ministry of Education, Science and Culture, Japan (24540283)
  and by Keio Gijuku Academic Development Funds.
  
\bibliographystyle{ws-procs975x65}
\bibliography{ws-pro-sample}

\end{document}